\documentclass[sn-basic]{sn-jnl}% Basic Springer Nature Reference Style/Chemistry Reference Style
\usepackage{mathtools}
\usepackage{blkarray, bigstrut}
\usepackage{graphicx}%
\usepackage{multirow}%
\usepackage{amsmath,amssymb,amsfonts}%
\usepackage{amsthm}%
\usepackage{mathrsfs}%
\usepackage[title]{appendix}%
\usepackage{xcolor}%
\usepackage{textcomp}%
\usepackage{manyfoot}%
\usepackage{booktabs}%
\usepackage{algorithm}%
\usepackage{algorithmicx}%
\usepackage{algpseudocode}%
\usepackage{listings}%
\theoremstyle{thmstyleone}%
\theoremstyle{thmstyletwo}%

\theoremstyle{thmstylethree}%
\newcommand{\parag}{\vspace{2ex}}

\begin{document}

\title[CVA Biplot based on the GSVD]{A Canonical Variate Analysis Biplot based on the Generalized Singular Value Decomposition}

\author*[1]{\fnm{Raeesa} \sur{Ganey}}\email{raeesa.ganey@wits.ac.za}

\author[2]{\fnm{Sugnet} \sur{Gardner-Lubbe}}\email{slubbe@sun.ac.za}

\affil[1]{\orgdiv{School of Statistics and Actuarial Science}, \orgname{University of Witwatersrand}, \orgaddress{\city{Johannesburg} \country{South Africa}}}

\affil[2]{\orgdiv{MuViSU (Center for Multi-dimensional Data Visualisation), Department of Statistics and Actuarial Science}, \orgname{Stellenbosch University}, \orgaddress{\city{Stellenbosch},  \country{South Africa}}}

\abstract{Canonical Variate Analysis (CVA) is a multivariate statistical technique and a direct application of Linear Discriminant Analysis (LDA) that aims to find linear combinations of variables that best differentiate between groups in a dataset. The data is partitioned into groups based on some predetermined criteria, and then linear combinations of the original variables are derived such that they maximize the separation between the groups. However, a common limitation of this optimization in CVA is that the within cluster scatter matrix must be nonsingular, which restricts the use of datasets when the number of variables is larger than the number of observations. By applying the generalized singular value decomposition (GSVD), the same goal of CVA can be achieved regardless on the number of variables. In this paper we use this approach to show that CVA can be applied and graphical representations to such data can be constructed. Specifically, we will be looking at the construction of a CVA biplot for such data that will display observations as points and variables as axes in a reduced dimension. Finally, we present experimental results that confirm the effectiveness of our approach.}


\maketitle

\section{Introduction}\label{Introduction}
\noindent Discriminant analysis is a predictive method that focuses on the relationship between a data matrix of predictor variables and a response variable. Its primary goal is to optimally separate different groups of objects to provide a classification rule for assigning entities of unknown origin to one of a known set of groups. Discriminant analysis helps researchers interpret the role of different predictor variables in group separability. In Linear Discriminant Analysis (LDA) \citep{fisher1936use}, researchers attempt to find linear combinations of the predictor variables that maximize the ratio of between-class variance to within-class variance. To do this, a pooled within-class covariance matrix is estimated, based on the assumption of equal within-class covariance matrices for the different groups. Canonical Variate Analysis (CVA) \citep{gittins1985canonical}, a direct extension of LDA, transforms the data into a reduced-dimensional space where cluster structure and class separability can be graphically examined. \parag

\noindent \citet{gabriel1971} introduced biplots as a graphical method for simultaneously displaying both samples and variables in multivariate data analysis, providing a comprehensive visualization tool that aids in understanding relationships and patterns within the data. The culmination of ideas on biplot methodology resulted in the publication of Gower and Hand's monograph in 1996 \citep{gower1996biplots}, which introduced a new perspective on the use of biplots as multivariate analogues of scatterplots. This approach allowed for the visualization of the underlying structure of high-dimensional data in a reduced number of dimensions, with biplot axes used to link the plotted points to the original variables in a manner similar to scatterplots. Gower and Hand noted that scatterplots were highly accessible, requiring little formal training to interpret, and emphasized that extending these principles to multivariate displays via biplots could enable non-statistical audiences to understand complex data more easily.  \parag

\noindent \cite{gabriel1972analysis} showed how to construct a CVA biplot that provides a two-dimensional graphical approximation of various groups optimally separated according to a multidimensional CVA criterion. By including details of group membership in the data matrix, researchers can construct biplots that are ideal for use in discriminant analysis. CVA biplots have proven useful in the context of classification problems, providing detailed insights into interclass structure, the role of different variables in overlap or separation between groups, and quantification of these factors. Graphical methods such as CVA biplots offer valuable insights beyond mere reporting of discriminant functions and error rates, making them a valuable tool for descriptive analysis of high-dimensional data.\parag

\noindent The goal of a CVA biplot is to discover a mapping that can transform each row of a data matrix ${\bf{X}}$ into a column within a lower-dimensional space, while exploring the cluster structure of the original data. To achieve this, the within-class and between-class scatter matrices are generated from ${\bf{X}}$ of size $n \times p$, and the traces of these matrices provide a means of quantifying the quality of the cluster relationship. Once the optimization criterion is defined in terms of these scatter matrices, the problem can be formulated as a generalized eigenvalue problem.  \parag

\noindent One of the main limitations of applying this approach of LDA is the case when the number of variables, $p$ should not exceed the number of observations, $n$, as LDA requires the within-class scatter matrix to be nonsingular. As \citet{ramey2016high} explain, researchers have emphasized standard approaches to modifying LDA for application to data where $p > n$, one of which includes covariance-matrix regularization. \parag

\noindent \citet{friedman1989regularized} introduced the well-known regularized discriminant analysis (RDA) classifier, which employs a biased covariance-matrix estimator. This estimator partially combines the sample covariance matrices obtained from linear and quadratic discriminant analysis and then shrinks the resulting estimator towards a scaled identity matrix. Numerous researchers have since then proposed further regularization techniques to enhance the estimation of group covariance matrices by adjusting the eigenvalues of sample covariance matrices to ensure positive definiteness. Some of these studies include \citet{ching2012regularized}, \citet{zhang2010regularized}, \citet{ji2008generalized}, \citet{guo2007regularized}, \citet{srivastava2007bayesian},  \citet{ye2006computational} and \citet{ye2005characterization}.  \parag 

\noindent Another important family of methods for dealing with singularity is based on a two-stage process in which two symmetric eigenproblems are solved successively. \citet{howland2003} looked at recasting the generalized eigenvalue problem in terms of a related generalized singular value problem, where they overcome this restriction on the relative dimensions of ${\bf{X}}$, therefore extending the applicability of LDA to any data matrix.  Their algorithm follows the generalized singular value decomposition (GSVD) which was first introduced by \citet{van1976generalizing} and then further formulated by \citet{edelman2020gsvd}. The GSVD of \citet{van1976generalizing} involves the decomposition of two matrices that are column-matched, and is different to the GSVD implemented in weighted solutions to Principal Component Analysis (PCA) and Correspondence Analysis. To date, no attempts have been made to construct a CVA biplot using solutions designed to overcome the limitations of LDA.  \parag

\noindent In this research paper, the work of \citet{howland2003} serves as a basis for extending the methodology to develop a CVA biplot that allows for unconstrained dimensionality of ${\bf{X}}$. The construction of biplots varies across different multivariate techniques, with each method requiring specific considerations based on the goal of the analysis. We provide a theoretical overview of the construction of the CVA biplot in Section 2, and delve into the GSVD in Section 3. Subsequent sections demonstrate the application of the GSVD in constructing the CVA biplot and experimental results with two sets of data. The CVA biplot constructed using the GSVD retains the primary goal of a traditional CVA biplot, which is to project the data into a reduced-dimensional space for the graphical exploration of cluster structure and class separability.\parag

\section{CVA and CVA Biplots}\label{CVA}
\noindent This section will explore the theoretical and practical implications of CVA for the analysis of high-dimensional data. CVA can be seen as an extension of Fisher's LDA \citep{fisher1936use} for data that can be classified into multiple groups. Fisher's original LDA aimed to identify a linear combination of $p$ measured variables that would optimally differentiate two groups. \citet{rao1948utilization} subsequently generalized Fisher's result to find a set of uncorrelated linear combinations that would best separate several groups. Notably, a greater degree of separation between groups leads to improved classification performance, making maximal group separation a crucial criterion for optimal discrimination. A useful measure of separation between groups is the ratio of a linear combination's total unconditional variance to its within-group variance, where a higher ratio indicates greater separation between groups. \parag

\noindent Consider the data matrix ${\bf{X}}$ consisting of $p$ variables measured on $n$ observations structured into $K$ groups, and $n > p$. We can represent the $i$-th observation belonging to the $k$-th group of ${\bf{X}}$ by the vector ${\bf{x}}_i^k$ of size $p$. The goal of CVA is to find a set of $p$ linear combinations of ${\bf{X}}$, ${\bf{l}}_1, {\bf{l}}_2, \dots, {\bf{l}}_p$, that maximize the between-group variation and minimize the within-group variation. These linear combinations are known as canonical variates. An important assumption of CVA is that the within-group covariances matrices of all groups are identical and important to check before performing CVA.  \parag

\noindent The between-group variation of ${\bf{X}}$ can be measured by the sum of squared deviations between the group means, denoted by

\begin{equation*}
\underset{p \times p}{\bf{B}} = \bar{{\bf{X}}}'{\bf{G}}'{\bf{G}}\bar{{\bf{X}}},
\end{equation*}

\noindent where ${\bf{G}}$: $n \times K$ denotes an indicator matrix defining the membership of the $K$ groups, and 

\begin{equation*}\label{xbar}
\bar{{\bf{X}}} = ({\bf{G}}'{\bf{G}})^{-1} {\bf{G}}'{\bf{X}}.
\end{equation*}

\noindent The within-group variation of ${\bf{X}}$ can be measured by the sum of squared deviations from their respective means, denoted by

\begin{equation*}\label{W}
\underset{p \times p}{\bf{W}} = {\bf{X}}' [ {\bf{I}} - {\bf{G}} ({\bf{G}}'{\bf{G}})^{-1} {\bf{G}}' ] {\bf{X}}.
\end{equation*}

\noindent The linear combination which optimally separates the observations from the different groups with respect to the $n$ samples is defined by the coefficient vector ${\bf{l}}$ which maximizes the variance ratio,

\begin{equation}\label{variance_ratio}
\lambda = \frac{{\bf{l}}' {\bf{B}}{\bf{l}}}{{\bf{l}}' {\bf{W}} {\bf{l}}}.
\end{equation}

\noindent The vector ${\bf{l}}$ that maximizes the variance ratio is only uniquely defined up to a scalar multiple. It is vital to impose a constraint on ${\bf{l}}$ that will confine the search space. A popular choice is the constraint that produces LDA exactly, 

\begin{equation}\label{constraint}
{\bf{l}}'{\bf{Wl}} = 1.
\end{equation}

\noindent The vector that maximizes the variance ratio in (\ref{variance_ratio}) and satisfies the constraint in (\ref{constraint}) is a non-zero solution to the two-sided eigenvalue problem:

\begin{align}
\frac{d}{d{\bf{l}}} \{  {\bf{l}}' {\bf{Bl}} - \lambda ({\bf{l}}'{\bf{Wl}} - 1 )  \} & = {\bf{0}}  \nonumber \\
2{\bf{Bl}} - 2 \lambda {\bf{Wl}} &= {\bf{0}}  \nonumber \\
{\bf{Bl}} &= \lambda {\bf{Wl}}.  \label{two-sided}
\end{align}

\noindent The solutions to the two-sided eigenvalue problem are the eigenvalues $\lambda_1, \lambda_2, \ldots, \lambda_p$ and eigenvectors ${\bf{l}}_1, {\bf{l}}_2, \ldots, {\bf{l}}_p$, since ${\bf{B}}$ is a $p \times p$ symmetric matrix and ${\bf{W}}$ is a $p \times p$ positive definite matrix (if $n < p$, then ${\bf{W}}$ is a positive semi-definite matrix). Thus, the largest eigenvalue $\lambda_1$ is the maximum value of the variance ratio and the coefficient vector that produces the maximum is the corresponding eigenvector ${\bf{l}}_1$. Often since the rank$({\bf{B}}) = $ min$(p,K-1)$, only ${\bf{l}}_1, \ldots, {\bf{l}}_{K-1}$ linear combinations are needed to maximize the between-group variation and minimize the within-group variation.\parag

\noindent \citet{brand2013pca} shows that the $p$ eigenvectors and eigenvalues can be simultaneously represented by 

\begin{equation*}\label{Two-sided}
{\bf{BL}} = {\bf{WL}}\pmb{\Lambda},
\end{equation*}

\noindent where $\pmb{\Lambda}$ is a $p \times p$ diagonal element matrix with $i$-th diagonal element equal to $\lambda_i$ and ${\bf{L}}$ is the $p \times p$ matrix with $i$-th column vector equal to ${\bf{l}}_i$, for $i = 1, 2, \ldots, p$. The $i$-th eigenvector of (\ref{two-sided}) maximises the ratio in (\ref{variance_ratio}) under the constraint in (\ref{constraint}) and conditional on being orthogonal to ${\bf{l}}_j$ in the metric {\bf{W}}, that is ${\bf{l}}'_i {\bf{W}}{\bf{l}}_j =0$, for all $ j < i $ \citep{rao1952advanced} $i = 2, \ldots, p$. The fact that the $p$ eigenvectors are orthogonal in ${\bf{W}}$ together with the set of constraints, implies that 

\begin{equation*}\label{constraint_b}
{\bf{L}}'{\bf{W}}{\bf{L}} = {\bf{I}}. 
\end{equation*}

\noindent \citet{brand2013pca} further shows that the eigenvalues in $\pmb{\Lambda}$ are identical to the eigenvalues of the matrix ${\bf{W}}^{-1/2}{\bf{BW}}^{-1/2}$, where ${\bf{W}}^{-1/2}$ is the square root of the inverse of the positive definite matrix ${\bf{W}}$. Furthermore, the eigenvectors that are orthonormal in the metric ${\bf{W}}$ are given by the column vectors of the matrix 

\begin{equation}\label{M}
\underset{p \times p}{\bf{L}} = {\bf{W}}^{-1/2}{\bf{V}},
\end{equation}

\noindent where ${\bf{V}}$ is the matrix with the $i$-th column vector given by the eigenvector of ${\bf{W}}^{-1/2}{\bf{BW}}^{-1/2}$ that corresponds to the $i$-th largest eigenvalue of ${\bf{W}}^{-1/2}{\bf{BW}}^{-1/2}$ and has unit length, $i = 1, 2, \ldots, p$. \parag

\noindent The observed values of the $p$ canonical variables for an observation with measurement vector ${\bf{x}}$ are given by the elements of the vector,

\begin{equation}\label{scores}
{\bf{y}}' = {\bf{x}}'{\bf{L}}.
\end{equation}

\noindent The original measured variables can be then be expressed as functions of the canonical variables:

\begin{equation}\label{vectors}
{\bf{x}}' = {\bf{y}}{'\bf{L}}^{-1}.
\end{equation} \parag

\noindent The CVA biplot is constructed using the scores on the first two canonical variables from CVA using (\ref{scores}) and (\ref{vectors}). These scores are used to plot observations from distinct groups and variables in a two-dimensional space. The position of each observation and variable in the biplot reflects their importance in discriminating between the groups. \parag 

\noindent In a CVA biplot, the observations belonging to different groups are represented as points. For any sample point ${\bf{x}}$, it is represented in the two-dimensional CVA biplot by the first two columns of ${\bf{x}}'{\bf{L}}$. Observations that are close together in the biplot are similar in their scores on the first two canonical variables, while observations that are far apart are dissimilar. \parag

\noindent The variables are represented as vectors, with the direction of each vector indicating the role of the variable in discriminating between the groups. Variables that are positively correlated with the first canonical variate will point in the same direction, while variables that are negatively correlated will point in the opposite direction. To construct vectors to represent the $p$ variables, we use (\ref{vectors}), which shows that the values of ${\bf{x}}$ may be predicted from the inner products of the sample point ${\bf{y}}$ in canonical space with the rows of ${\bf{L}}^{-1}$. \parag

\noindent Furthermore these vectors can be calibrated for prediction which refers to positions of marker points along the axes that accurately represent their corresponding values or measurements. The calibration allows for a meaningful interpretation of the distances between marker points along the axes. The position of the marker point for predicting a certain value $\mu$ on the $i$-th variable is given by 

\begin{equation*}
\frac{\mu}{{\bf{e}}_i' ({\bf{L}}^{-1})'{\bf{J}}{\bf{L}}^{-1}{\bf{e}}_i} {\bf{e}}_i' ({\bf{L}}^{-1})' {\bf{J}}, 
\end{equation*} 

\noindent where ${\bf{e}}_i$ is a column vector of which all elements are zero except for the $i$-th element which is equal to 1 and ${\bf{J}}$ is a $p \times p$ matrix with the first diagonal block being a $2 \times 2$ identity matrix reflecting the two-dimensional calibration and the other three blocks are zero matrices. \citet{gower2011understanding} provides further detail on how to calibrate the prediction axes.  \parag 

\noindent By examining the position of the variables and the different groups in the CVA biplot, we can gain insights into which variables are most important in discriminating between the groups, and which groups are most similar or dissimilar. This can help us to understand the underlying structure of the data and to make predictions about the group membership of new observations based on their values on the variables. There are several measures of cluster quality which involve the matrices ${\bf{W}}$ and ${\bf{B}}$. The following criterion will be used: 

\begin{equation}\label{J1}
trace({\bf{W}}^{-1}{\bf{B}}).
\end{equation}

 \parag 

\noindent A severe restriction to CVA and constructing its biplot, is that the within-group variation matrix ${\bf{W}}$ of a matrix ${\bf{X}}$ needs to be nonsingular, which restricts its application to datasets in which the number of variables $p$ exceed the number of observations $n$. The next sections covers the GSVD in which its application will allow users to construct a CVA biplot when $n < p$.

\section{The GSVD}

\noindent The GSVD is useful in many areas of mathematics and engineering, including system identification, control theory, signal processing, and statistics. One of its main advantages is that it allows us to study the relationship between two matrices in a systematic and structured way. Moreover, the GSVD can be used to solve many different problems, such as linear equations, least-squares problems, and eigenvalue problems. \parag

\noindent The GSVD is an extension of the classical Singular Value Decomposition (SVD) that allows the decomposition of two matrices simultaneously. It was originally defined by \citet{van1976generalizing} with some restrictions on the dimensions of the two matrices.  \citet{paige1981towards} generalised the decomposition to any matrices ${\bf{F}}  \in \mathbb{R}^{m_1 \times p}$ and ${\bf{H}}  \in \mathbb{R}^{m_2 \times p}$ and showed that the GSVD can be written as

\begin{equation}\label{GSVD_decomp}
\left [ \begin{array}{c}
{\bf{F}}\\
{\bf{H}}
\end{array} \right ] = \left [
\begin{array}{c c}
{\bf{UC}} \\
{\bf{VS}}
\end{array}
\right ] {\bf{M}}^{-1},
\end{equation}

\noindent where ${\bf{U}}$, ${\bf{V}}$ are square orthogonal matrices in $\mathbb{R}^{m_1 \times m_1}$, $\mathbb{R}^{m_2 \times m_2}$ respectively; ${\bf{C}} \in \mathbb{R}^{m_1 \times p}$ and ${\bf{S}}  \in \mathbb{R}^{m_2 \times p}$ are matrices defined below such that ${\bf{C}}'{\bf{C}} + {\bf{S}}'{\bf{S}} = {\bf{I}}_p $. Additionally, ${\bf{M}} \in \mathbb{R}^{p \times p}$ is a nonsingular matrix that relates the two decompositions and has rank $r$, where $r$ denotes the rank($[{\bf{F}};{\bf{H}}]$) \citep{edelman2020gsvd}. \parag

\begin{align*}
\mathbf{{\bf{C}} } &=
  \renewcommand{\arraystretch}{1.5}
  \begin{blockarray}{*{3}{c} c}
    \begin{block}{*{3}{>{$\footnotesize}c<{$}} c}
      $s$ & $r-s$ & $p-r$  \\
    \end{block}
    \begin{block}{[*{3}{c}]>{$\footnotesize}c<{$}}
    {\bf{I}} & {\bf{0}} & {\bf{0}} & $s$ \\
    {\bf{0}} & \text{diag}(\alpha_{s+1},\ldots,\alpha_r) & {\bf{0}} & $r-s$\\
    {\bf{0}} & {\bf{0}} & {\bf{0}} & $m_1 -r$ \\ 
    \end{block}
  \end{blockarray} \\
{\bf{S}} &= 
\renewcommand{\arraystretch}{1.5}
\begin{blockarray}{*{3}{c} c}
    \begin{block}{*{3}{>{$\footnotesize}c<{$}} c}
      $s$ & $r-s$ & $p-r$  \\
    \end{block}
    \begin{block}{[*{3}{c}]>{$\footnotesize}c<{$}}
    {\bf{0}} & {\bf{0}} & {\bf{0}} & $s$ \\
    {\bf{0}} & \text{diag}(\beta_{s+1},\ldots,\beta_r) & {\bf{0}} & $r-s$\\
    {\bf{0}} & {\bf{0}} & {\bf{I}} & $m_2 -r$ \\ 
    \end{block}
  \end{blockarray}
\end{align*}

\noindent where $s = r - \text{rank}({\bf{H}})$. The values $\alpha_i$ and $\beta_i$ are known as the singular values of the matrices ${\bf{F}}$ and ${\bf{H}}$, respectively and satisfy

\begin{equation*}
1 > \alpha_{s+1} \geq \ldots \geq \alpha_r > 0, \hspace{1.5cm} 0 < \beta_{s+1} \leq \ldots \leq \beta_r < 1
\end{equation*} 
and 
\begin{equation*}
\alpha_i^2 + \beta_i^2 = 1
\end{equation*} 
for $i = s+1, \ldots, r$. \parag

\noindent A further factorization illustrated in \citet{edelman2020gsvd} emphasizes the outer product rank $r$ form:

\begin{equation*}
\left [ \begin{array}{c}
{\bf{F}} \\
{\bf{H}} 
\end{array} \right ] = \sum_{i=1}^r \left [
\begin{array}{c}
\text{$i$-th column of} \left [
\begin{array}{c c}
{\bf{UC}} \\
{\bf{VS}}
\end{array}
\right ]
\end{array} \right ] \left [
\begin{array}{c}
\text{$i$-th row of ${\bf{M}}^{-1}$}
\end{array} \right ].
\end{equation*} \parag

\noindent \citet{paige1981towards} formulated the GSVD to include orthogonal matrices ${\bf{W}} \in \mathbb{R}^{r \times r}$ and ${\bf{Q}} \in \mathbb{R}^{r \times r}$  in addition to ${\bf{U}}$ and ${\bf{V}}$. The GSVD is defined such that it can be expressed as follows:
\begin{equation*}
{\bf{U}}'{\bf{F}}{\bf{Q}} = {\bf{C}}[ {\bf{W}}'{\bf{R}}, {\bf{0}} ]
\end{equation*}

\noindent and 

\begin{equation*}
{\bf{V}}'{\bf{H}}{\bf{Q}} = {\bf{S}}[{\bf{W}}'{\bf{R}}, {\bf{0}}]
\end{equation*}

\noindent and ${\bf{R}} \in \mathbb{R}^{r \times r} $ is nonsingular with its singular values equal to the nonzero singular values of ${\bf{K}}= [{\bf{F}};{\bf{H}}]$.

\noindent The proof of this formulation of the GSVD given in \citet{paige1981towards} begins with the complete orthogonal decomposition of $ {\bf{K}} $. 

\begin{equation*}
{\bf{P}}' {\bf{KQ}} = \left [
\begin{array}{c c}
{\bf{R}} & {\bf{0}}  \\
{\bf{0}} & {\bf{0}}
\end{array} \right ]
\end{equation*}

\noindent where ${\bf{P}}$ and ${\bf{Q}}$ are orthogonal and ${\bf{R}}$ is nonsingular with the same rank as ${\bf{K}}$. The construction proceeds by exploiting the SVDs of submatrices of ${\bf{P}}$. Partitioning ${\bf{P}}$ as 

\begin{equation*}
{\bf{P}} = \left [
\begin{array}{c c}
{\bf{P}}_{11} & {\bf{P}}_{12} \\ 
{\bf{P}}_{21} & {\bf{P}}_{22} 
\end{array} \right ]
\end{equation*} 

\noindent where ${\bf{P}}_{11} \in  \mathbb{R}^{m_1 \times r}$ and ${\bf{P}}_{21} \in  \mathbb{R}^{m_2 \times r}$ and thus implying $|{\bf{P}}_{11}|\leq 1$ which means that the singular values of ${\bf{P}}_{11}$ are less than one. The SVD can be written as ${\bf{U}}'{\bf{P}}_{11}{\bf{W}} = {\bf{C}}$. \parag

\noindent Next the decomposition ${\bf{P}}_{21}{\bf{W}} = {\bf{VL}}$, where ${\bf{V}} \in  \mathbb{R}^{p \times p}$ is orthogonal and ${\bf{L}}= (l_{ij})$ is the lower triangle with $l_{ij} = 0$ if $m_2 - i > r - j$ and $l_{ij} \geq 0$ if $m_2 - i = r - j$. This type of triangularization can be accomplished in the same way as QR decomposition \citep{gloub1996matrix} except that the columns are annihilated above the diagonal $m_2 - i = r - j$ working from right to left \citep{paige1981towards}. Then the matrix $[{\bf{C}};{\bf{L}}]$ has orthonormal columns, which implies that ${\bf{L}} = {\bf{S}}$. The results can be combined to complete the proof

\begin{align*}
\left [ \begin{array}{c}
{\bf{F}}\\
{\bf{H}}
\end{array} \right ] {\bf{Q}} &= \left [
\begin{array}{c c}
{\bf{P}}_{11} & {\bf{P}}_{12} \\ 
{\bf{P}}_{21} & {\bf{P}}_{22} 
\end{array} \right ] \left [ 
\begin{array}{c c}
{\bf{R}} & {\bf{0}}  \\
{\bf{0}} & {\bf{0}}
\end{array} \right ] \\ &=  \left [
\begin{array}{c c}
{\bf{P}}_{11}{\bf{R}} & {\bf{0}} \\ 
{\bf{P}}_{21}{\bf{R}} & {\bf{0}} 
\end{array} \right ]  \\ &=  \left [ 
\begin{array}{c c}
{\bf{U}}{\bf{C}}{\bf{W}}'{\bf{R}} & {\bf{0}}  \\
{\bf{V}}{\bf{S}}{\bf{W}}'{\bf{R}}  & {\bf{0}}
\end{array} \right ].
\end{align*} 

\noindent Therefore, 
%\begin{align}
%\left [ \begin{array}{c}
%{\bf{F}}\\
%{\bf{H}}
%\end{array} \right ]  &=   \left [ 
%\begin{array}{c}
%{\bf{U}}{\bf{C}} \\
%{\bf{V}}{\bf{S}}
%\end{array} \right ] \left [ 
%\begin{array}{c c}
%{\bf{W}}' {\bf{R}}  & {\bf{0}} \\
%\end{array} \right ] {\bf{Q}}'  \\ &= \left [ 
%\begin{array}{c}
%{\bf{U}}{\bf{C}} \\
%{\bf{V}}{\bf{S}}
%\end{array} \right ] {\bf{M}}^{-1},
%\end{align} 

\begin{align*}\label{MfromGSVD}
{\bf{M}}^{-1} = \left [ 
\begin{array}{c c}
{\bf{W}}'{\bf{R}} & {\bf{0}} \\
{\bf{0}} & {\bf{I}}_{p-r}
\end{array} \right ] {\bf{Q}}'.
\end{align*}

and 

\begin{align*}
{\bf{M}} = {\bf{Q}} \left [ 
\begin{array}{c c}
{\bf{R}}^{-1}{\bf{W}} & {\bf{0}} \\
{\bf{0}} & {\bf{I}}_{p-r}
\end{array} \right ].
\end{align*} \parag

\section{Application of the GSVD to CVA Biplots}\label{app_gsvd}
\noindent In this paper, we apply the GSVD method for constructing a CVA biplot in situations where the number of observations, $n$ is less than the number of variables, $p$, addressing the singularity issue of ${\bf{L}}$ in  (\ref{vectors}).   \parag

\subsection{GSVD in discriminant analysis}
\noindent We begin with the algorithm proposed by \cite{howland2003}, which for an $n \times p$ matrix ${\bf{X}}$ structured into distinct $K$ groups, the goal is to find a matrix ${\bf{G}}^*$: $ p\times (K-1)$ that preserves the cluster structure within a reduced-dimensional space. The algorithm uses ${\bf{G}}^*$ to compute the $(K-1)$-dimensional representation ${\bf{Y}}$ of ${\bf{X}}$. \parag

\noindent From equation (\ref{GSVD_decomp}) it follows that 

\begin{equation*}
{\bf{F}}'{\bf{F}} = ({\bf{M}}^{-1})'{\bf{C}}'{\bf{U}}'{\bf{U}}{\bf{C}}{\bf{M}}^{-1} = ({\bf{M}}^{-1})'{\bf{C}}'{\bf{C}}{\bf{M}}^{-1}
\end{equation*}

\noindent with 

\begin{align*}
\underset{p \times m_1 \times p}{{\bf{C}}'{\bf{C}}} &= \left[ 
  \renewcommand{\arraystretch}{1.5}
  \begin{array}{c c c}
    {\bf{I}}_s & {\bf{0}} & {\bf{0}} \\
    {\bf{0}} & \text{diag}(\alpha_{s+1}^2,\ldots,\alpha_r^2) & {\bf{0}} \\
    {\bf{0}} & {\bf{0}} & {\bf{0}} 
  \end{array} \right ],
 \end{align*}

\noindent where $m_1$ denotes the number of rows in ${\bf{F}}$, $r = rank([{\bf{F}};{\bf{H}}])$ and $s=r - rank({\bf{H}})$. Similary, from equation (\ref{GSVD_decomp}) it follows that 
 
 \begin{equation*}
 {\bf{H}}'{\bf{H}} = ({\bf{M}}^{-1})'{\bf{S}}'{\bf{V}}'{\bf{V}}{\bf{S}}{\bf{M}}^{-1} = ({\bf{M}}^{-1})'{\bf{S}}'{\bf{S}}{\bf{M}}^{-1}
 \end{equation*}
with

\begin{align*}
\underset{p \times m_2 \times p}{{\bf{S}}'{\bf{S}}} &= \left[ 
  \renewcommand{\arraystretch}{1.5}
  \begin{array}{c c c}
    {\bf{0}} & {\bf{0}} & {\bf{0}} \\
    {\bf{0}} & \text{diag}(\beta_{s+1}^2,\ldots,\beta_r^2) & {\bf{0}} \\
    {\bf{0}} & {\bf{0}} & {\bf{I}}_{m_2-r} 
  \end{array} \right ],
 \end{align*}

\noindent where $m_2$ denotes the number of rows in ${\bf{H}}$. In this manner, \citet{howland2003} showed that ${\bf{M}}^{-1}$ contains the right-generalized singular vectors of 

\begin{equation*}
\left [ 
\begin{array}{c}
{\bf{F}} \\ {\bf{H}} 
\end{array} \right ]: 2n \times p
\end{equation*}

\noindent where

\begin{equation*}
{\bf{F}} = {\bf{G}}\bar{\bf{X}}
\end{equation*}

\noindent and 

\begin{equation*}
{\bf{H}} = [ {\bf{I}} - {\bf{G}}({\bf{G}}'{\bf{G}})^{-1}{\bf{G}}' ] {\bf{X}}.
\end{equation*} \parag

\noindent The symmetric matrices ${\bf{B}}$ : $p \times p$ and ${\bf{W}}$ : $p \times p$ can be written as ${\bf{B}} = {\bf{F}}'{\bf{F}}$ and ${\bf{W}}= {\bf{H}}'{\bf{H}}$ where $m_1 = m_2 \equiv n$ such that 

\begin{equation*}
\left [ \begin{array}{c}
{\bf{B}} \\ {\bf{W}} 
\end{array} \right ] = \left [
\begin{array}{c}
{\bf{F}}'{\bf{F}} \\ {\bf{H}}'{\bf{H}}  
\end{array} \right ] = ({\bf{M}}^{-1})'  \left [
\begin{array}{c}
{\bf{C}}'{\bf{C}} \\ {\bf{S}}'{\bf{S}} 
\end{array} \right ] {\bf{M}}^{-1}.
\end{equation*} \parag

\noindent The GSVD can be written as  

\begin{equation*}
\left [ \begin{array}{c}
{\bf{F}}'{\bf{F}}\\
{\bf{H}}'{\bf{H}}
\end{array} \right ] {\bf{M}} = \left [
\begin{array}{c c}
({\bf{M}}^{-1})'{\bf{C}}'{\bf{C}} \\
({\bf{M}}^{-1})'{\bf{S}}'{\bf{S}} 
\end{array}
\right ].
\end{equation*}

\noindent Therefore,

 \begin{equation}\label{F'F}
 {\bf{F}}'{\bf{F}}{\bf{M}} = ({\bf{M}}^{-1})'{\bf{C}}'{\bf{C}},
 \end{equation}
 
 \noindent and 

\begin{equation}\label{H'H}
{\bf{H}}'{\bf{H}}{\bf{M}} = ({\bf{M}}^{-1})'{\bf{S}}'{\bf{S}}.
\end{equation} \parag

\noindent Writing ${\bf{m}}_{i}$ for the $i$-th column of ${\bf{M}}$ and ${\bf{m}}_{i}^{**}$ for the $i$-th row of ${\bf{M}}^{-1}$, it follows from equations (\ref{F'F})  and (\ref{H'H}),

\begin{equation*}
{\bf{F}}'{\bf{F}}{\bf{m}}_{i} = \alpha_i^2 {\bf{m}}_{i}^{**}
\end{equation*}

\noindent and 

\begin{equation*}
{\bf{H}}'{\bf{H}}{\bf{m}}_{i} = \beta_i^2 {\bf{m}}_{i}^{**},
\end{equation*}

\noindent from which we can write

\begin{equation*}
{\bf{m}}_{i}^{**} = \frac{1}{\alpha_i^2} {\bf{F}}'{\bf{F}}{\bf{m}}_{i} 
\end{equation*}
 and
  
\begin{equation*}
{\bf{m}}_{i}^{**} = \frac{1}{\beta_i^2} {\bf{H}}'{\bf{H}}{\bf{m}}_{i}
\end{equation*}

for $i = s+1, \ldots, r$. \parag

\noindent It follows that,

\begin{align*}\label{Bm=Wm}
\frac{1}{\alpha_i^2}{\bf{B}}{\bf{m}}_{i} &= \frac{1}{\beta_i^2}{\bf{W}}{\bf{m}}_{i} \\
{\bf{B}}{\bf{m}}_{i} & = \frac{\alpha_i^2}{\beta_i^2} {\bf{W}}{\bf{m}}_{i} \\
{\bf{B}}{\bf{m}}_{i} & = \lambda_i {\bf{W}}{\bf{m}}_{i}. 
\end{align*} \parag

\noindent where $\lambda_i = \frac{\alpha_i^2}{\beta_i^2}$. \parag

\noindent Therefore, the algorithm leverages the ${\bf{F}}$ and ${\bf{H}}$ matrices as inputs to the GSVD to obtain the decompositions, particularly focusing on the matrix ${\bf{M}}$. They include in ${\bf{G}}^*$ those columns in ${\bf{M}}$ which correspond to the $K-1$ largest $\lambda_i$'s, for $i = s+1, \ldots, r$. The $(K-1)$ representation ${\bf{Y}} = {\bf{X}} {\bf{G}}^*$ is used to define different maximization criterions in discriminant analysis when $p > n$ \citet{howland2003}. 
\parag

\subsection{GSVD to construct CVA biplots}

\noindent The algorithm of \cite{howland2003} leads to the proposed solution which uses the $s+1, \ldots, r$ columns of ${\bf{M}}$ to define a matrix 

\begin{equation*}
\underset{n \times q}{\bf{Z}} = {\bf{X}} \underset{p \times q}{\bf{M}},
\end{equation*}
to be used in the construction of a CVA biplot, where $q = r-s$.  Using ${\bf{Z}}$, a revised version of the  between and within-group variation matrices, ${\bf{B}}_{\text{gsvd}}$ and ${\bf{W}}_{\text{gsvd}}$ of size $q \times q$ are computed and are given by 

\begin{align}\label{B_gsvd}
{\bf{B}}_{\text{gsvd}} & = \bar{{\bf{Z}}}'{\bf{G}}'{\bf{G}}\bar{{\bf{Z}}},
\end{align}
\noindent and 
\begin{align}\label{W_gsvd}
{\bf{W}}_{\text{gsvd}} &= {\bf{Z}} '[ {\bf{I}} - {\bf{G}} ({\bf{G}}'{\bf{G}})^{-1} {\bf{G}}' ] {\bf{Z}}, 
\end{align}
where ${\bf{\bar{Z}}}$ represent the group means of ${\bf{Z}}$. \parag

\noindent With the GSVD approach, the within-group variation matrix, ${\bf{W}}_{\text{gsvd}}$ is nonsingular, enabling to compute

\begin{equation*}\label{M_gsvd}
{\bf{L}}_{\text{gsvd}} = {\bf{W}}_{\text{gsvd}}^{-1/2}{\bf{V}},
\end{equation*}

\noindent where ${\bf{V}}$ is the matrix with the $i$-th column vector given by the eigenvector of ${\bf{W}}_{\text{gsvd}}^{-1/2}{\bf{B}}_{\text{gsvd}}{\bf{W}}_{\text{gsvd}}^{-1/2}$ that corresponds to the $i$-th largest eigenvalue of ${\bf{W}}_{\text{gsvd}}^{-1/2}{\bf{B}}_{\text{gsvd}}{\bf{W}}_{\text{gsvd}}^{-1/2}$ and has unit length, $i = 1,2, \ldots, q$. \parag

\noindent The observed values of the canonical variables for an observation with a measurement vector ${\bf{x}}$ are represented by the elements of the vector 
 
\begin{equation}\label{samples_gsvd}
{\bf{y}}' = {\bf{z}}'{\bf{L}}_{\text{gsvd}} =  {\bf{x}}'{\bf{M}}{\bf{L}}_{\text{gsvd}}.
\end{equation}

\noindent The $p$ original measured variables can be expressed as a function of the canonical variables

\begin{equation*}\label{var_gsvd}
{\bf{x}}' = {\bf{y}}'({\bf{M}}{\bf{L}}_{\text{gsvd}})^{-1},
\end{equation*} \parag

\noindent where the inverse of the matrix ${\bf{M}}{\bf{L}}_{\text{gsvd}}$ is found using the Moore-Penrose generalized inverse  \citep{rakha2004moore} giving a unique result of a $q \times p$ matrix.  \parag

\noindent The position of the marker point for predicting a certain value of $\mu$ on the $i$-th variable is given by

\begin{equation}\label{var_c_gsvd}
\frac{\mu}{{\bf{e}}_i' (({\bf{M}}{\bf{L}}_{\text{gsvd}})^{-1})'{\bf{J}}({\bf{M}}{\bf{L}}_{\text{gsvd}})^{-1}{\bf{e}}_i} {\bf{e}}_i' (({\bf{M}}{\bf{L}}_{\text{gsvd}})^{-1})' {\bf{J}}.
\end{equation} \parag

\noindent  When constructing the CVA biplot through the GSVD, the number of observations represented by points remain $n$ from (\ref{samples_gsvd}) and the number of variables represented by the calibrated axes remain $p$ from (\ref{var_c_gsvd}), where $n$ and $p$ represent the original dimensions of ${\bf{X}}$. \parag

\noindent Furthermore, the cluster quality can now involve the matrices ${\bf{W}}_{\text{gsvd}}$ and ${\bf{B}}_{\text{gsvd}}$ similar to (\ref{J1}) which is given as 

\begin{equation}\label{J2}
trace({\bf{W}}_{\text{gsvd}}^{-1}{\bf{B}}_{\text{gsvd}}).
\end{equation} \parag

\noindent Unlike in PCA biplots, CVA biplots are invariant to the measurement scales  of the data used \citep{gower2011understanding}. Scaling the data does not impact the representation of samples and variables in a CVA biplot. However, when applying the GSVD approach, we have found that a scaling with a mean of zero and standard deviation of one on the columns of ${\bf{X}}$ needs to be implemented before inputting it into the GSVD algorithm. This step ensures that the resulting matrix ${\bf{M}}$, which is obtained through the GSVD algorithm used in turn to calculate ${\bf{Z}}$, will accurately represent the relationships and variations of the data in the biplot. 
The CVA biplot constructed using the GSVD will consistently display marker points on the calibrated axes in their original measurements. \parag

\noindent When $p < n $, the mathematical representation of samples as points in the standard CVA biplot and the biplot constructed through the GSVD approach is equivalent. Algebraically, this equivalence can be expressed as:

\begin{equation}\label{equal_samples}
{\bf{y}}'  = {\bf{x}}'{\bf{L}} =  {\bf{x}}'{\bf{M}}{\bf{L}}_{\text{gsvd}},
\end{equation}

\noindent which means that ${\bf{L}} = {\bf{M}} {\bf{L}}_{\text{gsvd}}$ and can be proven as follows: \parag

\noindent From equation (\ref{M}), ${\bf{L}} = {\bf{W}}^{-1/2} {\bf{V}}$ where ${\bf{V}}$ is an eigenvector of ${\bf{W}}^{-1/2}{\bf{B}} {\bf{W}} ^{-1/2}$, but ${\bf{L}}$ can also be expressed as an eigenvector of ${\bf{W}}^{-1}{\bf{B}}$. The spectral decomposition of ${\bf{W}}^{-1}{\bf{B}}$ is given as,   

\begin{equation}\label{sd1}
{\bf{V}}_1 \pmb{\Lambda}_1 {\bf{V}}_1', 
\end{equation}

\noindent where $ {\bf{V}}_1 = {\bf{L}} $. Similarly, the spectral decomposition of  ${\bf{W}}_{\text{gsvd}}^{-1}{\bf{B}}_{\text{gsvd}}$ is given as,

\begin{equation*}\label{sd2}
{\bf{V}}_2 \pmb{\Lambda}_2 {\bf{V}}_2'
\end{equation*}

\noindent where $ {\bf{V}}_2 = {\bf{L}}_{\text{gsvd}}$. \parag

\noindent From equations (\ref{B_gsvd}) and (\ref{W_gsvd}), ${\bf{B}}_{\text{gsvd}} = {\bf{M}}'{\bf{B}} {\bf{M}}$ and ${\bf{W}}_{\text{gsvd}} = {\bf{M}}'{\bf{W}} {\bf{M}}$, therefore,

\begin{align*}
{\bf{W}}_{\text{gsvd}}^{-1}{\bf{B}}_{\text{gsvd}} &= {\bf{M}}^{-1} {\bf{W}}^{-1} ({\bf{M}}^{-1})' {\bf{M}}' {\bf{B}} {\bf{M}}   \\
& = {\bf{M}}^{-1} {\bf{W}}^{-1} {\bf{B}} {\bf{M}} \\
& =  {\bf{M}}^{-1} {\bf{V}}_1 \pmb{\Lambda}_1 {\bf{V}}_1' {\bf{M}},
\end{align*}

\noindent Equating the above to equation (\ref{sd1}), 

\begin{equation*}
{\bf{V}}_2 \pmb{\Lambda}_2 {\bf{V}}_2' = {\bf{M}}^{-1} {\bf{V}}_1 \pmb{\Lambda}_1 {\bf{V}}_1' {\bf{M}}.
\end{equation*} \parag

\noindent This results in,
\begin{align*}
{\bf{V}}_2 &= {\bf{M}}^{-1} {\bf{V}}_1 \hspace{0.8cm} \text{and} \hspace{0.8cm}  \pmb{\Lambda}_2 = \pmb{\Lambda}_1\\
{\bf{V}}_1 &= {\bf{M}} {\bf{V}}_2, \\
{\bf{L}} &= {\bf{M}} {\bf{L}}_{\text{gsvd}}.
\end{align*} \parag

\noindent The representation of variables as axes is equivalent:

\begin{equation}\label{equal_var}
{\bf{x}}' =  {\bf{y}}{'\bf{L}}^{-1} = {\bf{y}}'({\bf{M}}{\bf{L}}_{\text{gsvd}})^{-1} 
\end{equation} 

\noindent Equations (\ref{equal_samples}) and (\ref{equal_var}) are only true when the variables in the data matrix ${\bf{X}}$ are standardized with mean of zero and standard deviation of one. \parag

\section{Experimental Results}
\noindent To compare the effectiveness of the standard CVA biplot construction method and the GSVD approach, two datasets where $n > p$ are first analyzed in Test I. Next, the CVA biplots constructed for datasets where $n < p$ using the GSVD approach are presented in Test II. The datasets used in both tests have group separations, and the resulting CVA biplots will showcase this feature and their corresponding cluster quality criterion are reported.  Additionally, due to the relatively large number of variables $p$ in Test II, only a selection of variables are displayed in the biplot. \parag

\subsection{Test I: $n>p$}

\noindent The \emph{Diabetes} dataset used in \citet{mclachlan2005discriminant} is a collection of medical data on 145 individuals, 85 of whom have been diagnosed with diabetes and 60 who have not. The subjects were classified based on conventional clinical criteria, into three groups: patients suffering from chemical diabetes, overt diabetes and normal subjects. The subjects were measured across three different variables: 

\begin{itemize}
\item \verb |glutest|: a measure of glucose intolerance.
\item \verb |instest|: a measure of insulin response to oral glucose.
\item \verb |sspg|: the steady-state plasma glucose concentration during an insulin suppression test.
\end{itemize}

\noindent \citet{mclachlan2005discriminant} used the \emph{Diabetes} dataset to illustrate the application of discriminant analysis and statistical pattern recognition techniques for predicting diabetes diagnosis based on a patient's medical profile. \parag

\noindent The \emph{Penguins} dataset is a collection of measurements on three species of penguins: Adelie, Gentoo, and Chinstrap. The dataset was collected by Dr. Kristen Gorman and the Palmer Station, Antarctica LTER, and first made available by the \textsf{R} software package \texttt{palmerpenguins} \citep{horst2022palmer}. The dataset contains the following variables for each of the 344 penguins:

\begin{itemize}
\item \verb |bl|: the length of the penguin's bill in millimeters.
\item \verb |bd|: the depth of the penguin's bill in millimeters.
\item \verb |fl|: the length of the penguin's flipper in millimeters.
\item \verb |bm|: Body mass (g): the body mass of the penguin in grams.
\end{itemize}

\noindent The \emph{Penguins} dataset has been widely used in data analysis and visualization research, as well as in teaching and learning activities. It has become a popular example of a multivariate dataset. \parag

\noindent  For both datasets, $n > p $, allowing us to construct both a standard CVA biplot and a biplot through the GSVD. The important assumption that the within-group covariance matrices of all groups are identical were checked prior to performing CVA on both datasets. To effectively compare the differences of the standard CVA biplot  and the biplot through the GSVD approach, the variables in the data are standardized (mean of zero and standard deviation of one). This is because the biplot constructed through the GSVD approach is sensitive to the scaling of the data, regardless that the standard CVA biplot is not.  \parag

\begin{figure}[ht]
\begin{center}
\includegraphics[width=0.8\columnwidth]{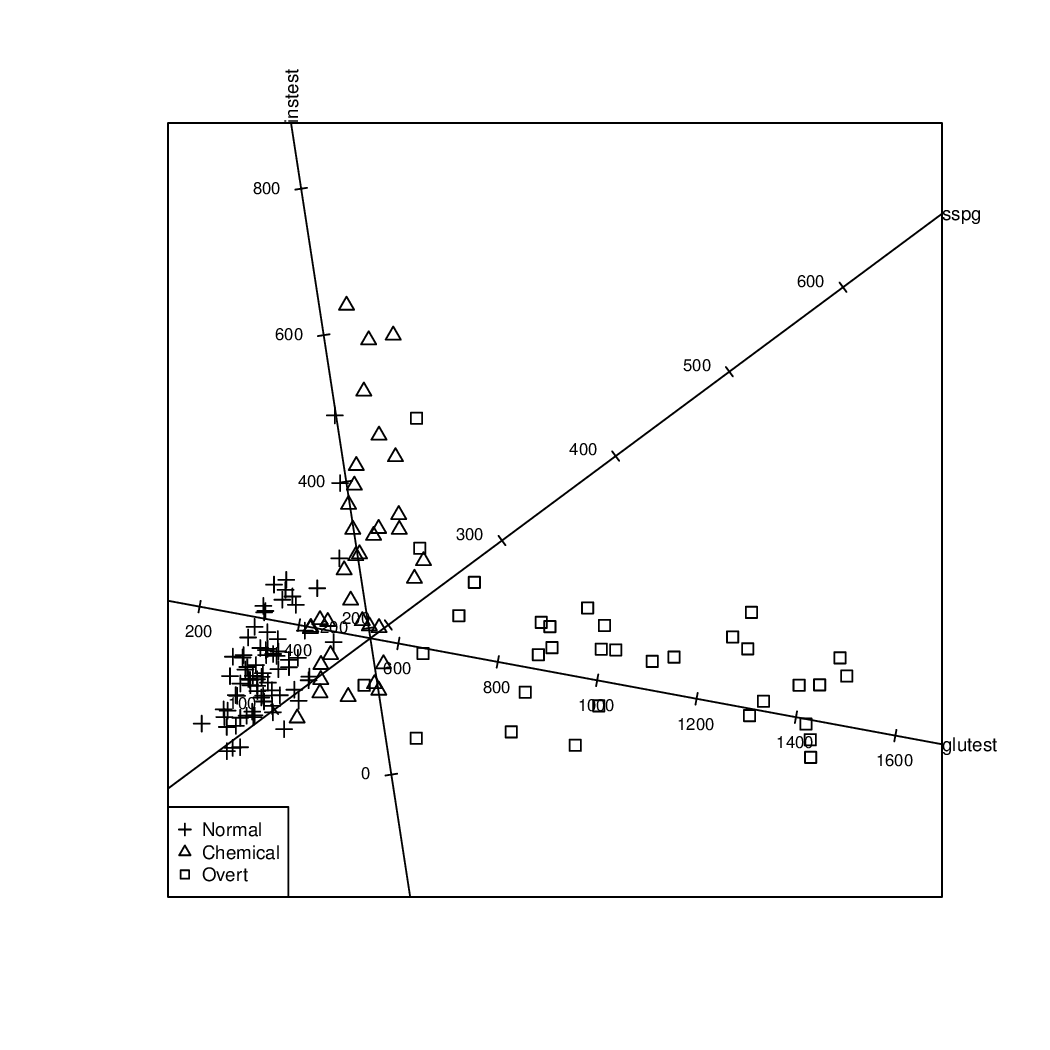}
  \caption{Resulting CVA Biplot of the \emph{Diabetes} dataset constructed through the standard method}
  \label{fig:diabetes}
  \end{center}
\end{figure}

\begin{figure}[ht]
\begin{center}
\includegraphics[width=0.8\columnwidth]{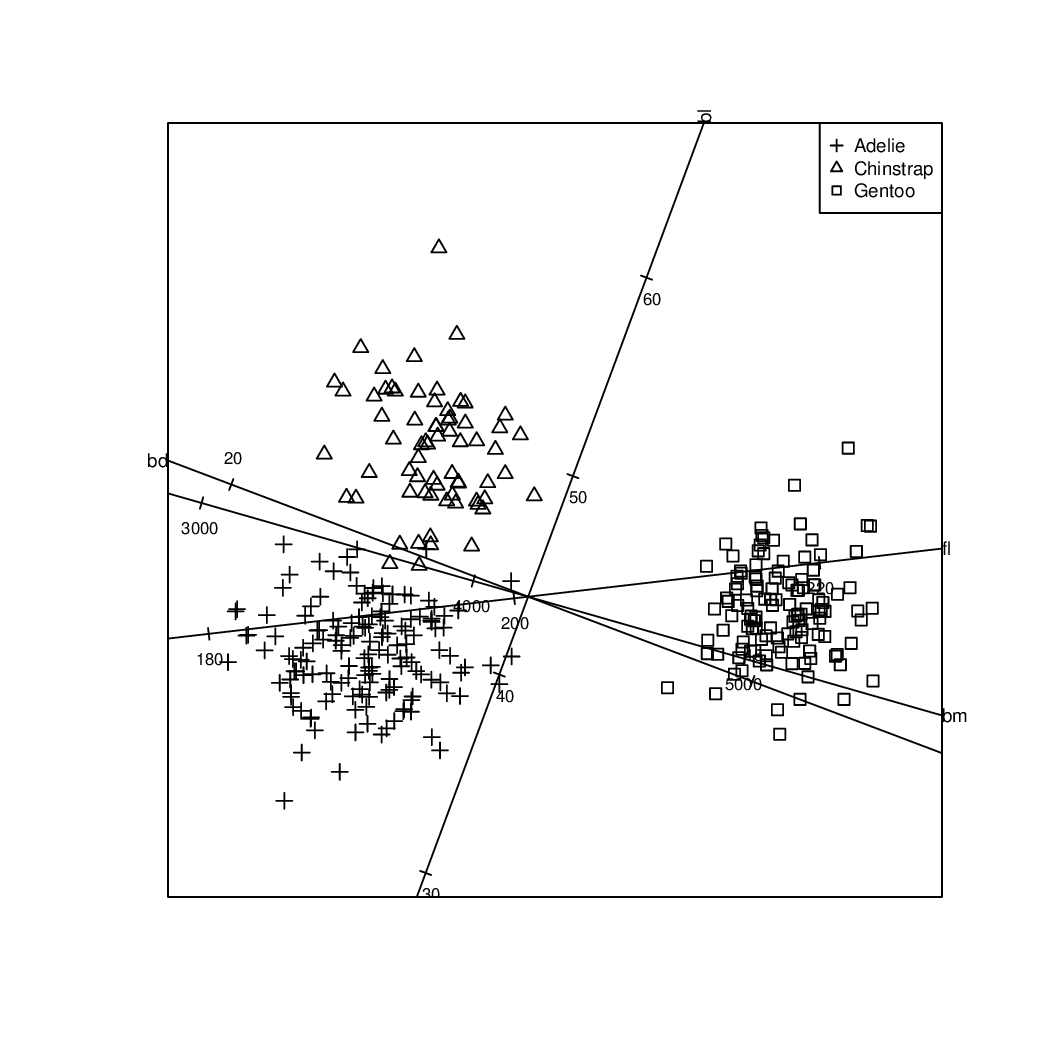}
  \caption{Resulting CVA Biplot of the \emph{Penguins} dataset constructed through the standard method}
  \label{fig:penguins}
    \end{center}
\end{figure}

\noindent The standard CVA biplot constructed for the \emph{Diabetes} dataset is shown in Figure \ref{fig:diabetes}, where the 145 patients are indicated by three different character points each belonging to the three groups, and the three variables by calibrated axes with marker points. The standard CVA biplot constructed for the \emph{Penguins} dataset is shown in Figure \ref{fig:penguins}, where the 344 penguins are indicated by three different character points each belonging to the three groups, and the four variables by calibrated axes with marker points.  \parag

\noindent The marker points on each axis in the biplot are given in the original scale of the respective variable. By maintaining the original scale of the variables, the marker points provide a meaningful reference for interpreting and understanding the relationship between the variables and the samples in the biplot. These marker points allow for a direct comparison with the observed data points, enabling insights into how specific variables contribute to the overall patterns and variations observed in the dataset. \parag

\begin{figure}[ht]
\begin{center}
\includegraphics[width=0.8\columnwidth]{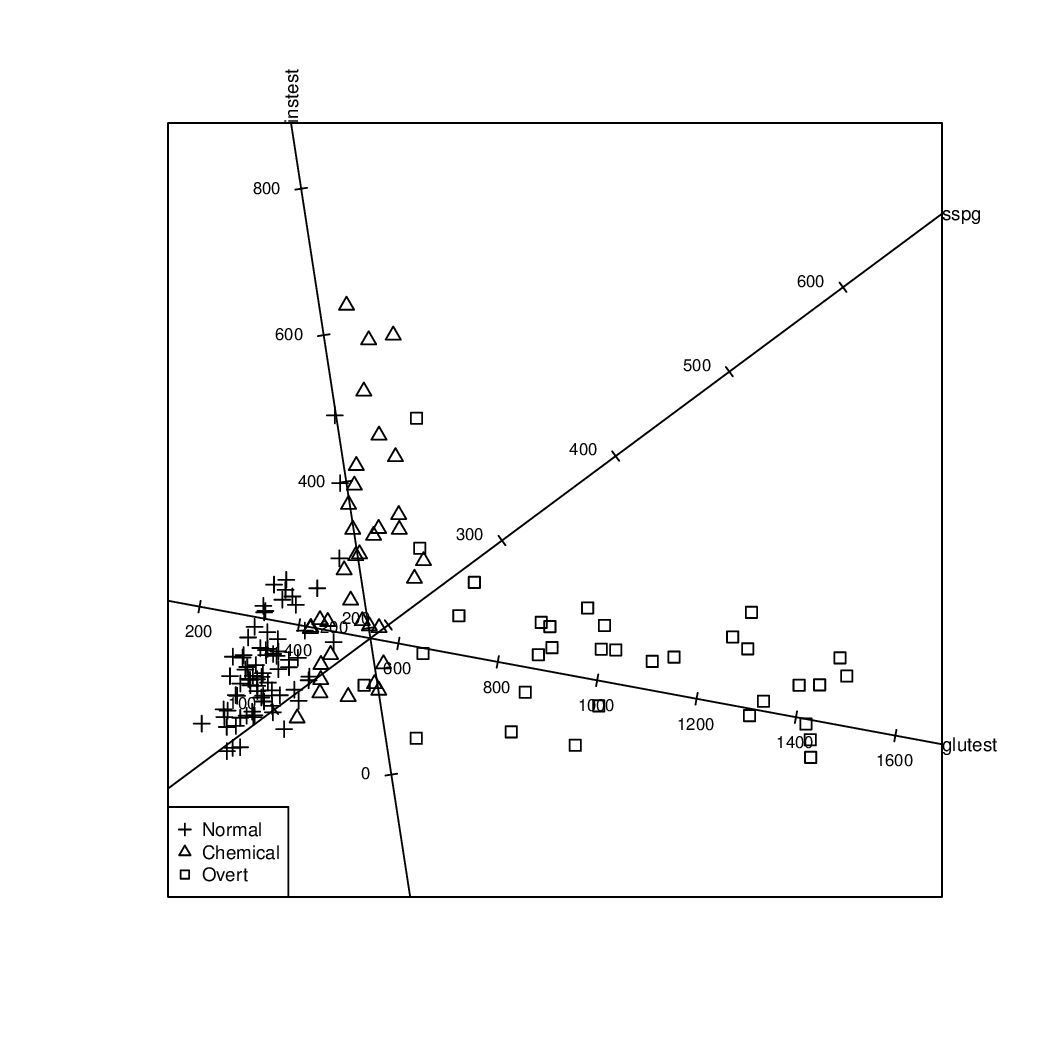}
  \caption{Resulting CVA Biplot of the \emph{Diabetes} dataset constructed through the GSVD approach}
  \label{fig:diabetesgsvd}
  \end{center}
\end{figure}

\begin{figure}[ht]
\begin{center}
\includegraphics[width=0.8\columnwidth]{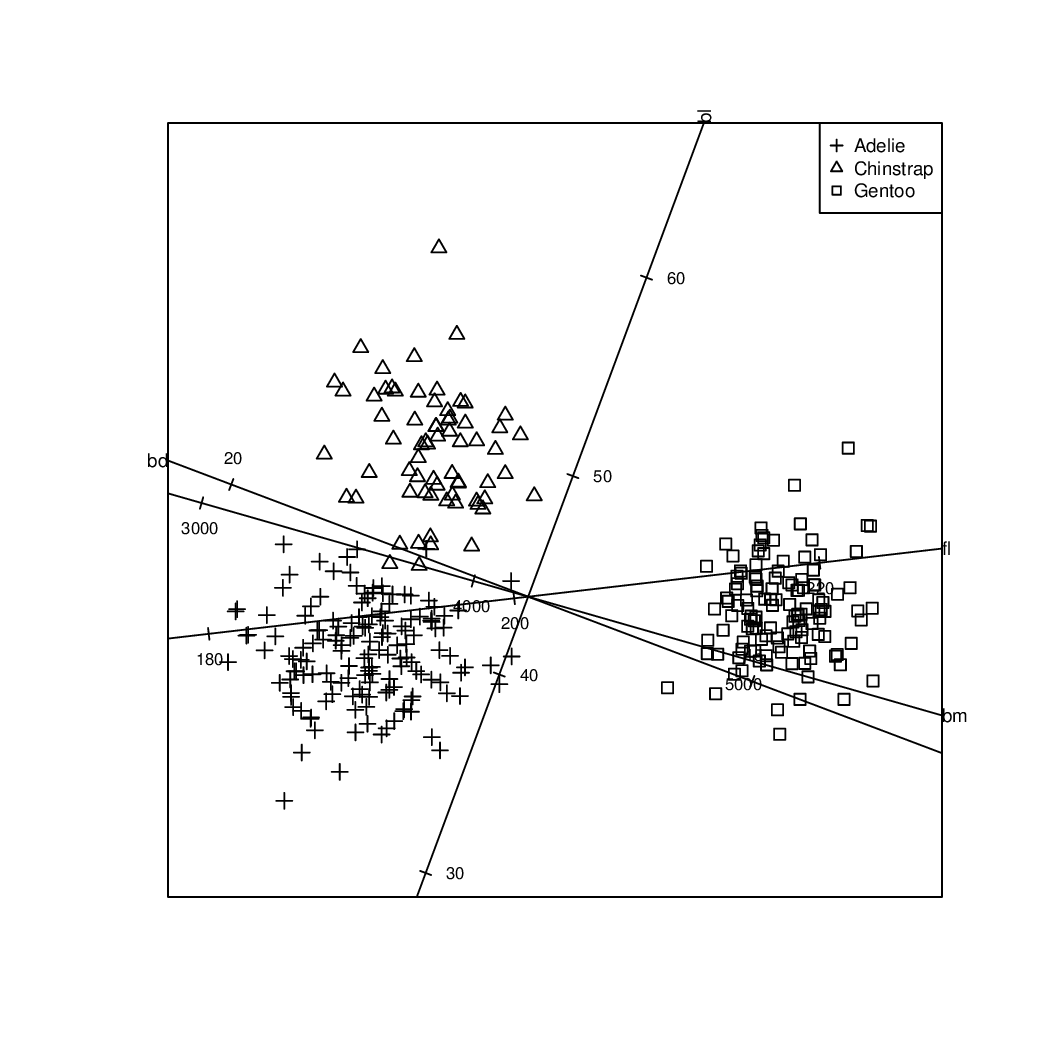}
  \caption{Resulting CVA Biplot of the \emph{Penguins} dataset constructed through the GSVD approach}
  \label{fig:penguinsgsvd}
    \end{center}
\end{figure}

\noindent The CVA biplot constructed through the GSVD produces an identical biplot as the standard CVA biplot for the \emph{Diabetes} and \emph{Penguins} datasets which is shown in Figures \ref{fig:diabetesgsvd} and \ref{fig:penguinsgsvd}. The cluster quality from (\ref{J1}) and (\ref{J2}), which refers to the sums of the diagonal elements of the matrix obtained by taking the inverse of the within-group variation matrix and then multiplying it with the matrix of between-group variation also produces the same value. This is given in Table \ref{cluster_quality} for both datasets. \parag 

\noindent The CVA biplots obtained from both datasets, shown in Figures \ref{fig:diabetes} and \ref{fig:penguins}, along with the cluster quality evaluations presented in Table \ref{cluster_quality}, yield consistent results for both the standard method and the GSVD approach. This outcome implies that when constructing CVA biplots with $n < p$, the GSVD approach converges to the standard method. Therefore, in situations where the within-group variation matrix ${\bf{W}}$ is singular, the GSVD approach serves as a reliable alternative to the standard method. \parag

\begin{table}[!h]
\caption{Comparison of cluster quality from the standard CVA biplot construction and the GSVD approach for data where $n > p$.}\label{cluster_quality}
\setlength{\tabcolsep}{12.2pt} % Default value: 6pt
\renewcommand{\arraystretch}{1.4} % Default value: 1
\begin{tabular}{| c | c | c | c | c | c |}
\hline 
Data & $n$ & $p$ & $K$ & $trace({\bf{W}}^{-1}{\bf{B}})$ & $trace({\bf{W}}_{\text{gsvd}}^{-1}{\bf{B}}_{\text{gsvd}})$ \\ \hline \hline
\emph{Diabetes} & 145 & 4 & 3 & 4.1156 & 4.1156 \\ \hline 
\emph{Penguins} & 344 & 4 & 3 & 17.3422 & 17.3422 \\ \hline
\end{tabular}
\end{table}

\subsection{Test II: $n < p $}
\noindent Two cancer datasets are used to illustrate the CVA biplot through the GSVD when the number of variables is larger than the number of observations. The first data, \emph{NCI60} processed by \citet{culhane2003cross} is a dataset of gene expression profiles of 60 National Cancer Institute (NCI) cell lines. The 60 human tumour cell lines are derived from patients with leukaemia, melanoma, along with, lung, colon, central nervous system (CNS), ovarian, renal, breast and prostate cancers. The second data, \emph{Colon cancer} gene expression data obtained from \citet{alon1999broad} and studied by \citet{shafi2020detection}, are made up of 62 cases and 2000 genes from patients with colon cancer. Patients' tumor biopsies are grouped either as Abnormal or Normal.  Table \ref{cluster_quality2} displays the values of $n$, $p$ and $K$ for these datasets. \parag 

\noindent In these type of datasets, the number of variables exceeds the number of observations, which limits the possibility of constructing a CVA biplot solely through the GSVD approach. The procedure to construct such a biplot remains the same, but it is important to be mindful of the computational challenges associated with handling a high-dimensional dataset including the testing of the CVA assumption. \parag

\begin{figure}[ht]
\begin{center}
\includegraphics[width=0.8\columnwidth]{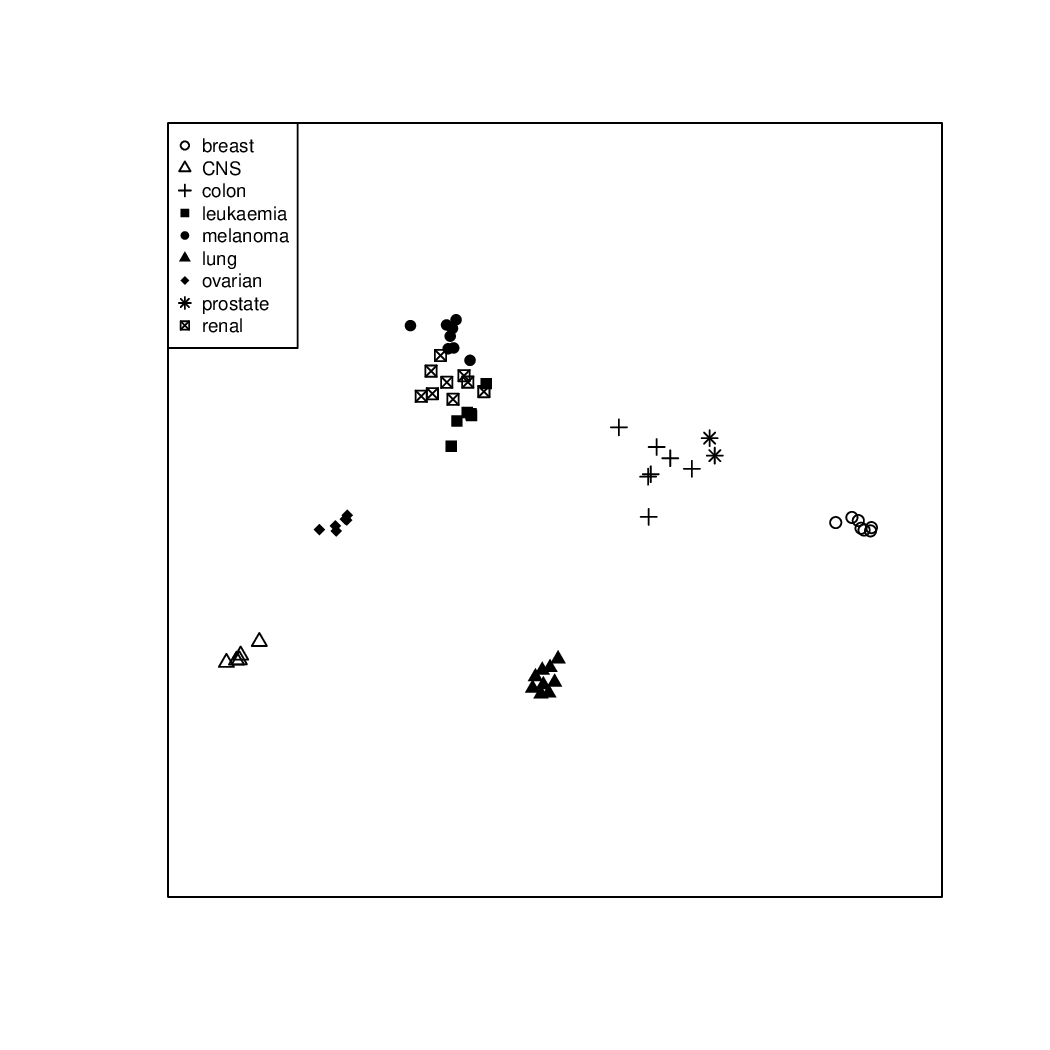}
  \caption{Resulting CVA display of observations in the \emph{NCI60} dataset constructed through the GSVD approach}
  \label{fig:nci}
    \end{center}
\end{figure}

\begin{figure}[ht]
\begin{center}
\includegraphics[width=0.80\columnwidth]{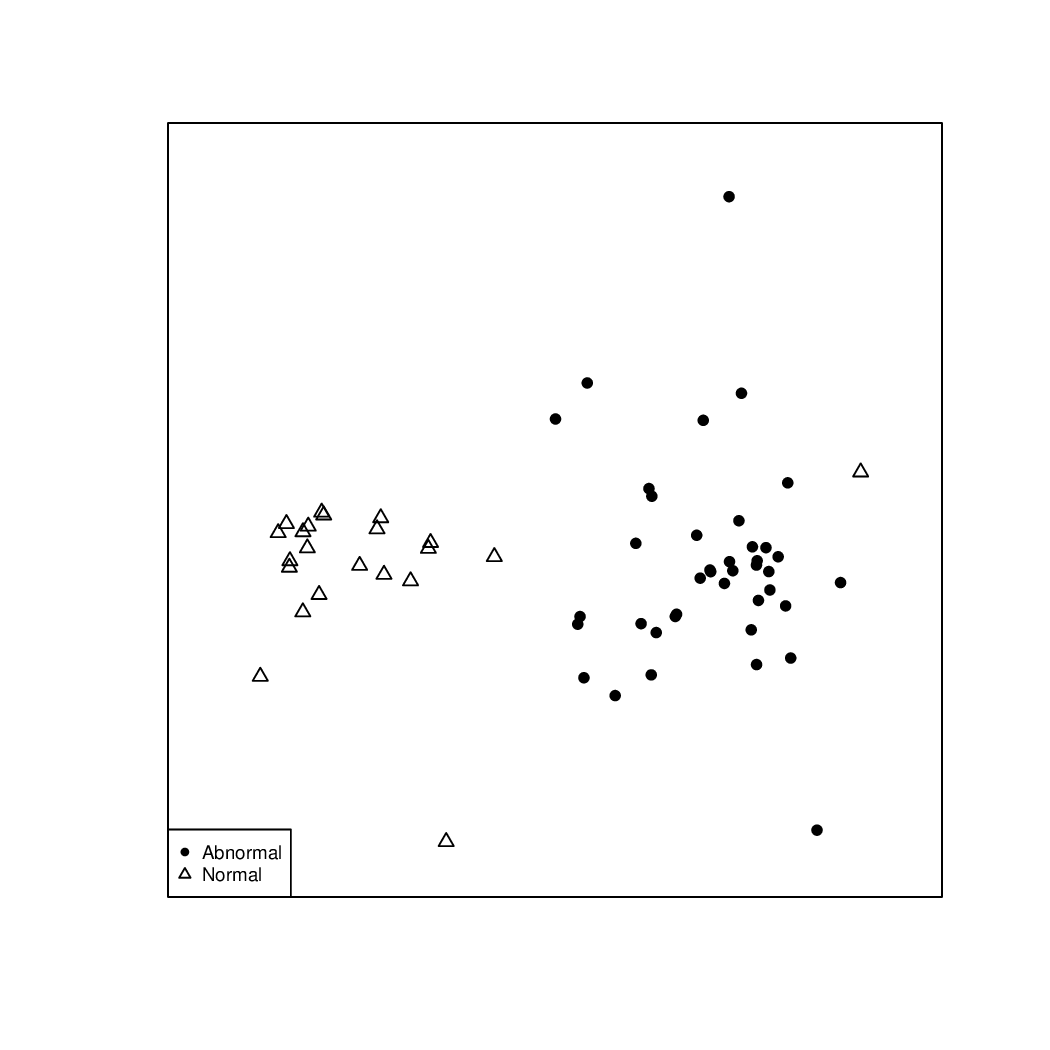}
  \caption{Resulting CVA display of observations in the \emph{Colon cancer} dataset constructed through the GSVD approach}
  \label{fig:colon}
    \end{center}
\end{figure}

\noindent As $p$ is relatively large and it would therefore be difficult to visualize all the variables as axes in a CVA biplot. Given this, only the observations from the \emph{NCI60} and \emph{Colon cancer} datasets are shown in the CVA displays in Figures \ref{fig:nci} and \ref{fig:colon}, respectively. \parag

\noindent Observations in Figure  \ref{fig:nci} belonging to the 9 different groups in the \emph{NCI60} data are easily distinguishable in the CVA display.
Examining closely one can see there are three groups close together, namely: melanoma, renal and leukaemia. The colon and prostate cancer groups are also closely related. \parag

\noindent In the CVA display of the \emph{Colon cancer} data in Figure \ref{fig:colon}, individual observations are depicted. Grouping of observations can be easily distinguished, except for one major anomaly. A tumor biopsy classified as Normal lies in the Abnormal group of tumor biopsies. \parag

\noindent In situations where $n < p$, the cluster quality measure in (\ref{J1}) cannot be applied due to the singularity of the within-group variation matrix ${\bf{W}}$, and therefore the cluster quality measure in (\ref{J2}) is used. A higher ratio indicates better cluster quality. Table \ref{cluster_quality2} shows displays the values for each dataset.  \parag

\begin{table}[!h]
\caption{Cluster quality GSVD approach for data where $n <p$.}\label{cluster_quality2}
\setlength{\tabcolsep}{12.2pt} % Default value: 6pt
\renewcommand{\arraystretch}{1.4} % Default value: 1
\begin{tabular}{| c | c | c | c | c |}
\hline 
Data & $n$ & $p$ & $K$ & $trace({\bf{W}}_{\text{gsvd}}^{-1}{\bf{B}}_{\text{gsvd}})$ \\ \hline \hline
%\emph{Colon Cancer} & 62 & 2000 & 2 & 0.0625 & 0.0417 \\ \hline 
\emph{NCI60} & 60 & 6830 & 9 & 4.615  \\ \hline 
\emph{Colon} & 62 & 2000 & 2 &  3.484   \\ \hline
%\emph{Simulated} & 100 & 300 & 4 & 0.0444 & 0.0656  \\ \hline 
\end{tabular}
\end{table}

\newpage

\subsection{Displaying variables in a CVA biplot when $p$ is large}
\noindent When $n>p$ in Test I, analyzing the biplots enables one to determine the key variables responsible for distinguishing between groups and to identify the most distinct groups. On the other hand, in Test II datasets where $p$ is often considerably large, visualizing all $p$ variables in a single CVA biplot may pose challenges. Two main strategies are proposed to address this issue: 

\begin{enumerate}
\item Variable selection: Prioritize important or relevants variables for the analysis based on domain knowledge or specific research objectives. The researcher can select a subset of variables that contribute most to the separation between groups or have the highest variation. This can help reduce the number of variables in the biplot. 
\item Interactive visualization: Utilize interactive feactures in visualization tools or software that allow researchers to explore the biplot and dynamically display or hide variables as needed. This approach provides flexibility in examining specific variables or subsets while maintaining an overall view. 
\end{enumerate} \parag 

\begin{figure}[ht]
\begin{center}
\includegraphics[width=0.80\columnwidth]{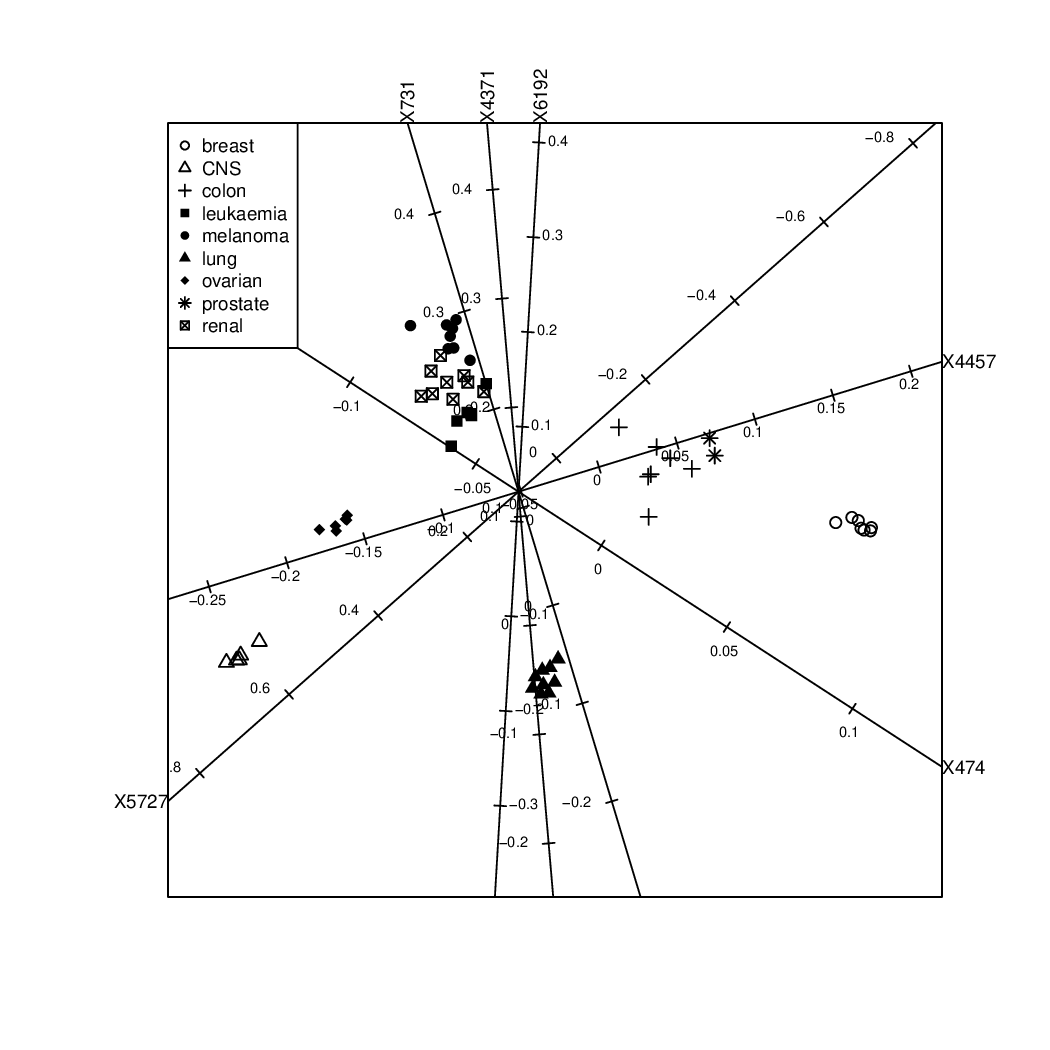}
  \caption{Resulting CVA Biplot of the \emph{NCI60} dataset constructed through the GSVD approach with selection of six gene variables.}
  \label{fig:nci2}
    \end{center}
\end{figure}

\noindent For the purposes of this paper, the first strategy is applied to the \emph{NCI60} cancer dataset in Test II. For the \emph{NCI60} dataset, six gene variables are selected to be displayed in the CVA biplot of Figure \ref{fig:nci2}. The biplot with these set of axes allows one to see that there are two-three main variable groups that contribute to the separation of the observations. Only six variables were chosen to avoid overcrowding the biplot with axes and to highlight the potential such a visualization.  \parag

\noindent The inclusion of marker points along the axes in a biplot significantly enhances the ability to identify and interpret the predicted values of various variables, particularly when analyzing outliers or anomalies within the dataset. These markers act as reference points, making it easier to understand the distribution of values and their relationship to other variables. For instance, when analyzing genetic data, a researcher may focus on specific genes that are relevant to distinct groups within the dataset. By referencing the marker points on the axes, they can determine the range of values associated with each group, enabling a more detailed understanding of group-specific characteristics and trends. This approach provides clarity in distinguishing patterns and aids in identifying key variables that contribute to the differences between groups. \parag

\section{Conclusion}
\noindent In conclusion, biplots have emerged as a valuable tool for visualizing high-dimensional data and providing insights into complex relationships between variables. Discriminant analysis and CVA biplots have been particularly useful for classification problems, revealing the interclass structure and the role of different variables in overlap or separation between groups. However, the limitation of the approach of LDA, which restricts the number of variables to not exceed the number of observations makes it difficult to construct a CVA biplot. \parag

\noindent Despite numerous endeavors documented in the literature aimed at mitigating the limitations of LDA, including RDA and the contributions made by \citet{howland2003}, a significant gap persists in the availability of visualization techniques tailored to these enhanced methods. Traditional visualization methods, such as biplots, have not been extended to incorporate the advancements made in LDA variants. However, the GSVD algorithm proposed in this paper offers a promising solution to this challenge. By allowing for the unconstrained dimensionality of the data matrix, the GSVD enables the construction of CVA biplots for any dataset, irrespective of the size of $p$. \parag

\noindent Biplots are recognized as powerful and accessible tools for the analysis of high-dimensional data, providing intuitive visual representations that facilitate interpretation and insight generation. Through this demonstration, the paper aimed to showcase the potential of GSVD-based CVA biplots as a valuable analytical tool for researchers working with such datasets, offering enhanced interpretability and insight into multivariate data structures. \parag

\newpage

\section{Data Availability}
The datasets utilized in this study are publicly accessible through specific R packages as detailed below:
\begin{enumerate}
\item \emph{Diabetes}: This dataset can be accessed through the \emph{heplots} R package \citep{heplots}. 
\item \emph{Penguins}: This dataset is available via the \emph{palmerpenguins} R package \citep{palmerpenguins}.
\item \emph{NCI60}: This dataset is accessible through the \emph{ISLR} R package \citep{islr}.
\item \emph{Colon cancer}: This data can be accessed through the \emph{plsgenomics} R package \citep{plsgenomics}. 
\end{enumerate}

\bibliography{Manuscript}% common bib file

\begin{thebibliography}{33}
\providecommand{\natexlab}[1]{#1}
\providecommand{\url}[1]{{#1}}
\providecommand{\urlprefix}{URL }
\providecommand{\doi}[1]{\url{https://doi.org/#1}}
\providecommand{\eprint}[2][]{\url{#2}}
 \bibcommenthead

\bibitem[{Alon et~al(1999)Alon, Barkai, Notterman, Gish, Ybarra, Mack, and
  Levine}]{alon1999broad}
Alon U, Barkai N, Notterman DA, et~al (1999) Broad patterns of gene expression
  revealed by clustering analysis of tumor and normal colon tissues probed by
  oligonucleotide arrays. Proceedings of the National Academy of Sciences
  96(12):6745--6750

\bibitem[{Boulesteix et~al(2024)Boulesteix, Durif, Lambert-Lacroix, Peyre, and
  Strimmer.}]{plsgenomics}
Boulesteix AL, Durif G, Lambert-Lacroix S, et~al (2024) plsgenomics: PLS
  Analyses for Genomics.
  \urlprefix\url{https://CRAN.R-project.org/package=plsgenomics}

\bibitem[{Brand(2013)}]{brand2013pca}
Brand H (2013) {PCA} and {CVA} biplots: A study of their underlying theory and
  quality measures. PhD thesis, Stellenbosch: Stellenbosch University

\bibitem[{Ching et~al(2012)Ching, Chu, Liao, and Wang}]{ching2012regularized}
Ching WK, Chu D, Liao LZ, et~al (2012) Regularized orthogonal linear
  discriminant analysis. Pattern Recognition 45(7):2719--2732

\bibitem[{Culhane et~al(2003)Culhane, Perri{\`e}re, and
  Higgins}]{culhane2003cross}
Culhane AC, Perri{\`e}re G, Higgins DG (2003) Cross-platform comparison and
  visualisation of gene expression data using co-inertia analysis. BMC
  bioinformatics 4(1):1--15

\bibitem[{Edelman and Wang(2020)}]{edelman2020gsvd}
Edelman A, Wang Y (2020) The {GSVD}: {W}here are the ellipses?, matrix
  trigonometry, and more. SIAM Journal on Matrix Analysis and Applications
  41(4):1826--1856

\bibitem[{Fisher(1936)}]{fisher1936use}
Fisher RA (1936) The use of multiple measurements in taxonomic problems. Annals
  of eugenics 7(2):179--188

\bibitem[{Friedman(1989)}]{friedman1989regularized}
Friedman JH (1989) Regularized discriminant analysis. Journal of the American
  statistical association 84(405):165--175

\bibitem[{Friendly et~al(2024)Friendly, Fox, and Monette}]{heplots}
Friendly M, Fox J, Monette G (2024) {heplots}: Visualizing Tests in
  Multivariate Linear Models.
  \urlprefix\url{https://CRAN.R-project.org/package=heplots}

\bibitem[{Gabriel(1971)}]{gabriel1971}
Gabriel KR (1971) The biplot graphic display of matrices with application to
  principal component analysis. Biometrika 58(3):453--467

\bibitem[{Gabriel(1972)}]{gabriel1972analysis}
Gabriel KR (1972) Analysis of meteorological data by means of canonical
  decomposition and biplots. Journal of Applied Meteorology (1962-1982) pp
  1071--1077

\bibitem[{Gittins(1985)}]{gittins1985canonical}
Gittins R (1985) Canonical variate analysis. Canonical Analysis: A Review with
  Applications in Ecology pp 67--95

\bibitem[{Gloub and Van~Loan(1996)}]{gloub1996matrix}
Gloub GH, Van~Loan CF (1996) Matrix computations. Johns Hopkins Universtiy
  Press, 3rd edtion

\bibitem[{Gower and Hand(1996)}]{gower1996biplots}
Gower JC, Hand DJ (1996) Biplots, vol~54. CRC Press

\bibitem[{Gower et~al(2011)Gower, Lubbe, and Le~Roux}]{gower2011understanding}
Gower JC, Lubbe SG, Le~Roux NJ (2011) Understanding {B}iplots. John Wiley \&
  Sons

\bibitem[{Guo et~al(2007)Guo, Hastie, and Tibshirani}]{guo2007regularized}
Guo Y, Hastie T, Tibshirani R (2007) Regularized linear discriminant analysis
  and its application in microarrays. Biostatistics 8(1):86--100

\bibitem[{Horst et~al(2020)Horst, Hill, and Gorman}]{palmerpenguins}
Horst AM, Hill AP, Gorman KB (2020) palmerpenguins: Palmer Archipelago
  (Antarctica) penguin data. \doi{10.5281/zenodo.3960218},
  \urlprefix\url{https://allisonhorst.github.io/palmerpenguins/}

\bibitem[{Horst et~al(2022)Horst, Hill, and Gorman}]{horst2022palmer}
Horst AM, Hill AP, Gorman KB (2022) Palmer archipelago penguins data in the
  palmerpenguins {R} {P}ackage-{A}n {A}lternative to {A}nderson’s {I}rises. R
  JOURNAL 14(1):244--254

\bibitem[{Howland et~al(2003)Howland, Jeon, and Park}]{howland2003}
Howland P, Jeon M, Park H (2003) Structure preserving dimension reduction for
  clustered text data based on the generalized singular value decomposition.
  SIAM Journal on Matrix Analysis and Applications 25(1):165--179

\bibitem[{James et~al(2021)James, Witten, Hastie, and Tibshirani}]{islr}
James G, Witten D, Hastie T, et~al (2021) ISLR: Data for an Introduction to
  Statistical Learning with Applications in R.
  \urlprefix\url{https://CRAN.R-project.org/package=ISLR}

\bibitem[{Ji and Ye(2008)}]{ji2008generalized}
Ji S, Ye J (2008) Generalized linear discriminant analysis: a unified framework
  and efficient model selection. IEEE Transactions on Neural Networks
  19(10):1768--1782

\bibitem[{McLachlan(2005)}]{mclachlan2005discriminant}
McLachlan GJ (2005) Discriminant analysis and statistical pattern recognition.
  John Wiley \& Sons

\bibitem[{Paige and Saunders(1981)}]{paige1981towards}
Paige CC, Saunders MA (1981) Towards a generalized singular value
  decomposition. SIAM Journal on Numerical Analysis 18(3):398--405

\bibitem[{Rakha(2004)}]{rakha2004moore}
Rakha MA (2004) On the {M}oore--{P}enrose generalized inverse matrix. Applied
  Mathematics and Computation 158(1):185--200

\bibitem[{Ramey et~al(2016)Ramey, Stein, Young, and Young}]{ramey2016high}
Ramey JA, Stein CK, Young PD, et~al (2016) High-dimensional regularized
  discriminant analysis. arXiv preprint arXiv:160201182

\bibitem[{Rao(1948)}]{rao1948utilization}
Rao CR (1948) The utilization of multiple measurements in problems of
  biological classification. Journal of the Royal Statistical Society Series B
  (Methodological) 10(2):159--203

\bibitem[{Rao(1952)}]{rao1952advanced}
Rao CR (1952) Advanced statistical methods in biometric research.

\bibitem[{Shafi et~al(2020)Shafi, Molla, Jui, and Rahman}]{shafi2020detection}
Shafi A, Molla MI, Jui JJ, et~al (2020) Detection of colon cancer based on
  microarray dataset using machine learning as a feature selection and
  classification techniques. SN Applied Sciences 2:1--8

\bibitem[{Srivastava et~al(2007)Srivastava, Gupta, and
  Frigyik}]{srivastava2007bayesian}
Srivastava S, Gupta MR, Frigyik BA (2007) Bayesian quadratic discriminant
  analysis. Journal of Machine Learning Research 8(6)

\bibitem[{Van~Loan(1976)}]{van1976generalizing}
Van~Loan CF (1976) Generalizing the singular value decomposition. SIAM Journal
  on numerical Analysis 13(1):76--83

\bibitem[{Ye and Yu(2005)}]{ye2005characterization}
Ye J, Yu B (2005) Characterization of a family of algorithms for generalized
  discriminant analysis on undersampled problems. Journal of Machine Learning
  Research 6(4)

\bibitem[{Ye et~al(2006)Ye, Xiong, and Madigan}]{ye2006computational}
Ye J, Xiong T, Madigan D (2006) Computational and theoretical analysis of null
  space and orthogonal linear discriminant analysis. Journal of Machine
  Learning Research 7(7)

\bibitem[{Zhang et~al(2010)Zhang, Dai, Xu, and Jordan}]{zhang2010regularized}
Zhang Z, Dai G, Xu C, et~al (2010) Regularized discriminant analysis, ridge
  regression and beyond. The Journal of Machine Learning Research 11:2199--2228

\end{thebibliography}
%% if required, the content of .bbl file can be included here once bbl is generated
%%\input sn-article.bbl

\end{document}